\documentclass[twocolumn,english,floatfix,aps,prl]{revtex4}
\usepackage[T1]{fontenc}
\usepackage[latin1]{inputenc}

\makeatletter

\usepackage{babel}
\usepackage{graphicx}
\makeatother
\begin{document}

\preprint{This line only printed with preprint option}

\title{High Field behavior of the Spin Gap compound Sr$_2$Cu(BO$_3$)$_2$}

\author{Suchitra E. Sebastian$^1$, D. Yin$^1$, P.
Tanedo$^1$, G. A. Jorge$^1$, N. Harrison$^2$, M. Jaime$^2$, Y.
Mozharivskyj$^3$, G. Miller$^3$, J. Krzystek$^4$, S. A.
Zvyagin$^4$, I. R. Fisher$^1$}

\affiliation{$^1$Geballe Laboratory for Advanced Materials and
Department of Applied Physics, Stanford University, Stanford, CA
94305}

\affiliation{$^2$MST-NHMFL, Los Alamos National Laboratory, Los
Alamos, NM 87545}

\affiliation{$^3$Department of Chemistry, Iowa State University,
Ames, IA 50011}

\affiliation{$^4$National High Magnetic Field Laboratory,
Tallahassee, FL 32310}

\begin{abstract}
We report magnetization and heat capacity measurements of 
single crystal samples of the spin gap compound Sr$_2$Cu(BO$_3$)$_2$. 
Low-field data show that the material has a singlet ground state comprising dimers with intradimer 
coupling J = 100 K. High field data reveal the role of weak interdimer coupling. For fields that are large compared to the spin gap, 
triplet excitations are observed for significantly smaller fields 
than predicted for isolated dimers, 
indicating that weak inter-dimer coupling leads to triplet delocalization. 
High field magnetization 
behavior at low temperatures suggests additional cooperative effects.
\end{abstract}


\date{\today}

\maketitle

Recently, the spin gap system SrCu$_2$(BO$_3$)$_2$ which is a
physical realization of the Shastry-Sutherland model [1], has
received a considerable amount of interest [2-5]. Spin frustration
in a network of connected dimers leads to competition between Neel
and singlet dimer ground states, reducing the spin gap to $\sim$ 30
K. The excited state magnetization is found to be a series of
fractional magnetic plateaus of the total magnetization,
associated with localized triplet excitations in structures
commensurate with the lattice. In this letter, we report results
on the first single crystals of the spin gap system
Sr$_2$Cu(BO$_3$)$_2$ which although similar in composition to
SrCu$_2$(BO$_3$)$_2$, has a very different magnetic lattice and
exhibits novel magnetic properties associated with triplet
delocalization.

Sr$_2$Cu(BO$_3$)$_2$ exists in two structural phases: here we
describe the first measurements of single crystal samples of the
high temperature $\beta$-Sr$_2$Cu(BO$_3$)$_2$ phase. This phase is
thermodynamically stable above $\sim $ 800 $^{\circ} $C, so samples
were quenched from 860 $ ^{\circ} $C to room temperature during
the growth process. $\beta$-Sr$_2$Cu(BO$_3$)$_2$  has an
orthorhombic unit cell with lattice parameters \emph{a} =
7.612\AA, \emph{b} = 10.854\AA and \emph{c} = 13.503\AA [6]. The
structure is two-dimensional, comprising layers of Cu$_2$B$_4$O$
_{12}$ containing Cu $^{2+}$ (3d$^9$, s = $\frac{1}{2}$) ions (shown in Figure 1), separated by Sr ions [6]. Each layer comprises distorted
octahedral Cu(1)O$_6$ units, square planar Cu(2)O$_4$ units and
triangular B(1,2,3)O$_3$ units.

\begin{figure}[htbp]
\includegraphics[width=0.43\textwidth]{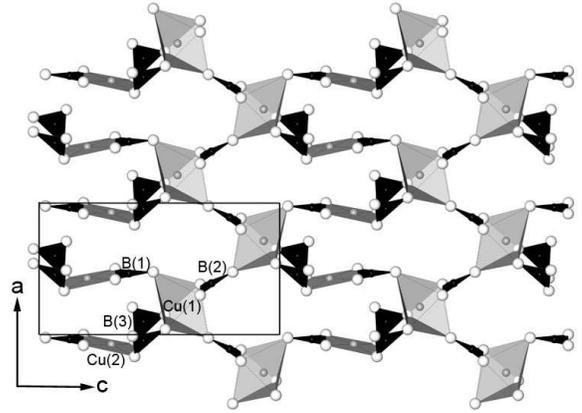}
\caption{Cu$_2$B$_4$O$ _{12}$ layers from
$\beta$-Sr$_2$Cu(BO$_3$)$_2$ . The open circles represent O, the
gray circles Cu and the black circles B. Dimer units comprise
Cu(1) and Cu(2) atoms linked via B(3)O$_3$ groups. Sr ions
separating the layers are not shown.}
\end{figure}

The Cu$^{2+}$ ions are coupled via three different exchange
pathways (shown as J, J' and J'' in Figure 2) via triangular
BO$_3$ units. Orthorhombic distortion of the Cu(1)O$_6$
octahedra lifts the e$_g$ degeneracy of this ion 
(apical oxygen atoms are
further from the central Cu(1) atom at 2.49\AA and 2.42\AA, than
the equatorial oxygen atoms, at 1.99\AA and 1.92\AA [6,7]), such
that the highest energy level is the antibonding combination of Cu
3d$_{x^2-y^2}$ and the equatorial O 2p$_\sigma$ orbitals. Since the
Cu(1) 3d$^9$ hole resides in the equatorial plane, the 
dominant exchange pathway is therefore J, through the equatorial
oxygen ions of the Cu(1)O$_6$ octahedra. The equivalent magnetic lattice (Figure 2a) then comprises Cu(1)-Cu(2) dimers coupled by J via B(3)O$_3$ triangular units. In this paper, we present high field magnetization measurements that reveal the effect of weak inter-dimer coupling J' and J" through B(2)O$_3$ and B(1)O$_3$ groups respectively.

\begin{figure}[htbp]
\includegraphics[width=0.5\textwidth]{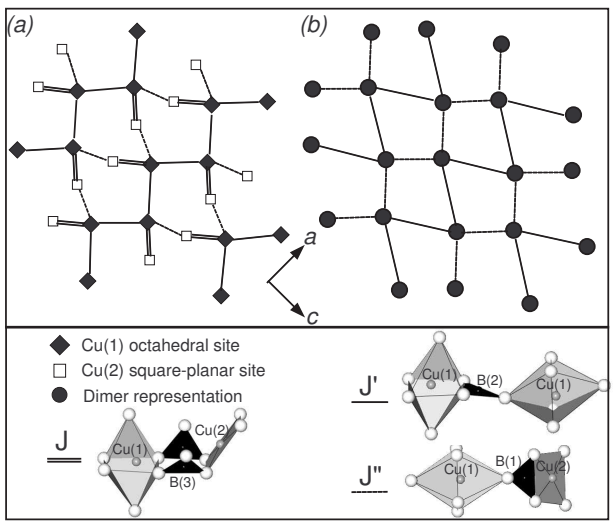}
\caption{(a) Magnetic lattice representing the structure of
Cu$_2$B$_4$O$ _{12}$ planes in Sr$ _2$Cu(BO$_3$)$_2$. 
Atomic positions of Cu(1) and Cu(2) are taken from the real crystal structure. 
Exchange pathways are labelled in the legend. 
(b) In the limit J $\gg$ J' and J'', interdimer terms can be treated perturbatively, and the lattice mapped on to a distorted square lattice of dimers. 
Each dimer unit is represented by a solid circle  
positioned in the geometric center of the Cu(1) and Cu(2) atoms. 
}
\end{figure}

Single crystals of $\beta$-Sr$_2$Cu(BO$_3$)$_2$ were grown using
LiBO$_2$ as a flux [6]. The polycrystalline precursor was prepared
by a solid state reaction of SrCO$_3$, CuO and B$_2$O$_3$ ground
together in stoichiometric ratios and heated in flowing O$_2$
 at 900$^{\circ}$ C for approximately 72 hours with intermediate
grindings. A mixture with 1:1.4 molar ratio of polycrystalline
Sr$_2$Cu(BO$_3$)$_2$  to LiBO$_2$ was heated in a Pt crucible to
925$^{\circ}$C, cooled at approximately 1$^{\circ}$C per hr to
860$^{\circ}$C and the remaining liquid decanted. The crystals
were allowed to cool rapidly in air. The resulting single crystals
of $\beta$-Sr$_2$Cu(BO$_3$)$_2$ were multi-faceted, purple in
color, with dimensions of up to 5mm on a side.

Magnetization measurements in fields up to 5 T were performed in a Quantum Design SQuID magnetometer. DC magnetic
susceptibility ($\chi$ = $\frac{M}{H}$) measured in 5000 Oe with H aligned
along the principal crystal axes is shown in Figure 3. From the
exponential behavior at low temperatures, it is clear that the
material is a spin gap system. To estimate the intradimer
coupling J, a simple approximation is to fit the magnetic
susceptibility to an isolated dimer model with an additional Curie
term to account for a small isolated impurity spin concentration:

\[
\chi =
\frac{N(\mu{}_B{g})^2}{k{}_BT(3+\exp(\frac{J}{k_BT}))}+\frac{C}{T}+\chi{}_0\]
where N is Avogadro's number, C the Curie constant due to
non-interacting impurities and $\chi$$_0$ a
temperature-independent term. Any renormalization due to
interdimer coupling terms would modify the value of g. For all
three orientations, the magnetic susceptibility per mole of Cu
spins as a function of temperature can be well fit by the isolated
dimer model. The value of J for all three orientations is found
from these fits to be 100 $\pm$ 1.5 K in agreement with previous
polycrystalline data [7]. Values for g are found to be 2.0 $\pm$
0.05 for H aligned along the \textit{b} and \textit{c}-axes, but
2.2 $\pm$ 0.05 for H aligned along the \textit{a}-axis, consistent
with the orientation of the CuO$_4$ units. C varies from 0.001 to
0.003 emuK/molOe, corresponding to 0.3$\%$ to 0.8$\%$ impurity
concentration, assuming impurity spins with g = 2, s = $\frac{1}{2}$. The
T-independent term $\chi$$_0$ is small and positive, likely due to
a very small concentration of ferromagnetic impurities. Values
vary slightly between samples and with orientation, and for the
data shown in Figure 3 correspond to approximately 10$^{-4}$
$\mu${}$_B$ per formula unit.

\begin{figure}[htbp]
\includegraphics[width=0.49\textwidth]{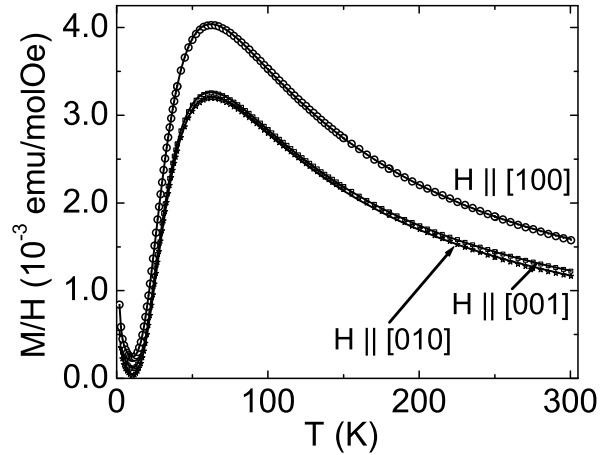}
\caption{Temperature dependence of the magnetic susceptibility of
a single crystal of Sr$_2$Cu(BO$_3$)$_2$ for a field of 5000 Oe
aligned along the principal crystal axes. Solid lines show fit to
isolated dimer model, described in main text.}
\end{figure}

Independent measurements of g are necessary to determine the extent of renormalization of the magnetic susceptibility due to interdimer coupling. Values for g were obtained via Electron Spin
Resonance (ESR) using a Bruker Elexsys E680X spectrometer at X-band frequency 9.38 GHz at room temperature. These measurements gave g$_\emph{a} = 2.230$, g$_\emph{b} = 2.060$ and g$_\emph{c} =2.130$ $\pm$ 0.004 in good agreement with values obtained from susceptibility fits within the uncertainty of these measurements, indicating that any interdimer terms are not large enough to have a measurable effect at low fields.

An additional confirmation of the value of J in this system is
obtained from heat capacity measurements. The Schottky anomaly
from the large spin gap is too small to be easily resolved against
the large phonon contribution to the heat capacity. However, the
shift in heat capacity due to Zeeman splitting of the triplet
states in a magnetic field can be observed. Measurements were made
in zero field and 18 T continuous field for a single crystal of
$\beta$-Sr$_2$Cu(BO$_3$)$_2$ weighing 7.9 mg. The field was
oriented at an arbitrary angle to the crystal. A calorimeter made
of plastic materials and silicon was used, employing a thermal
relaxation time technique. The difference between these two
values, $\Delta$C$_p$ = C$_p$(18T) - C$_p$(0T) is due solely to
changes in the magnetic contribution and can be calculated for an
isolated dimer model using measured values for J and g for this
crystal. Figure 4 shows calculated and measured values for
$\Delta$C$_p$ as a function of temperature. The measured data has
a broad maximum with a peak value of 0.38 $\pm$ 0.07 J/molK
centered at 14 $\pm$ 1 K, compared to the model which has a peak
value of 0.32 J/molK centered at 16 K. The data agree within
experimental uncertainty, confirming that the isolated dimer model
with J = 100 K is a reasonable first approximation in small
magnetic fields.

We note that additional features are evident in the heat capacity
at high temperatures (inset to figure 4), which appear to be
unrelated to the low temperature magnetic properties. Structural
refinements performed on a single crystal indicate no change in
symmetry or average atomic positions between the Low Temperature
(LT) (T $<$ 230 K), Room Temperature (RT) (230 K $<$ T $<$ 320 K), and
High Temperature (HT) (T $>$ 320 K) regions, but a marked change in
the thermal parameters of the B atoms indicates a change in
vibrational properties.

\begin{figure}[htbp]
\includegraphics[width=0.49\textwidth]{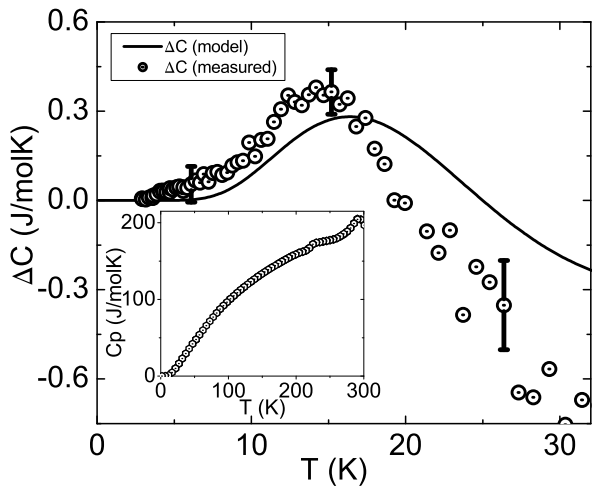}
\caption{Difference in heat capacity $\Delta$C$_p$ = C$_p$(18 T) -
C$_p$(0 T) for a single crystal of Sr$_2$Cu(BO$_3$)$_2$. Solid
line shows calculated value for dimer model with J = 100 K and g =
2.14. Inset shows total heat capacity in zero field between 2 K
and 300 K. Mol refers to the formula unit.}
\end{figure}

The behavior of Sr$_2$Cu(BO$_3$)$_2$ is most interesting in high
magnetic fields, for which g$\mu$$_B$H $\sim$ J. In particular, we
find that the magnetization is no longer fit well by the isolated
dimer model, due to the effects of interdimer coupling.
Magnetization measurements were made in pulsed high magnetic
fields up to 65 T at different temperatures. The data are obtained
using a wire-wound sample extraction magnetometer in which the
sample is inserted or removed from the detection coils \emph{in
situ}. Data were taken for increasing and decreasing fields, and
the average value calculated. A collection of six randomly
oriented fragments from a clean single crystal was used in order
to maximize the filling factor of the coil. Magnetization for this
composite sample as a function of field is shown in Figure 5. A
small linear diamagnetic contribution with negative slope
9x10$^{-4}$ $\mu${}$_B$/T, due to the response of the source
coils, has been subtracted from the data. The value of
magnetization at 50 K in a field of 5 T was compared with SQuID
magnetometer magnetization data in order to obtain absolute values
for the magnetization in units of $\mu${}$_B$/Cu.

A feature of the isolated dimer model is a sharp increase in
magnetization to 1 $\mu${}$_B$ when the lowest Zeeman-split
triplet state crosses the singlet state at T = 0. For J = 100 K
and g = 2.14 (i.e. average value for the composite sample), this
would occur at an applied magnetic field H$_c$ = 69 T if there was
no interdimer interaction, with thermal broadening at finite
temperatures (solid lines in Figure 5). As shown in Figure 5, M(H)
data correspond very closely to this model for temperatures of 20
K and above, but deviate significantly for lower temperatures. At
1.5 K, data taken to 65 T clearly show the swift rise in
magnetization due to triplet excitations. This occurs at a
magnetic field lower than J. Furthermore, data for temperatures of
5 and 10 K coincide with the 1.5 K data within experimental
uncertainty.

The deviation from the isolated dimer model is clearly revealed in
a graph of magnetization as a function of temperature (Figure 6).
In high magnetic fields, the magnetization does not fall to zero at low temperatures, but is as large as 0.15 $\mu${}$_B$
for temperatures below 10 K. Above 15 K, the M(T) curves rise
sharply, coinciding with the isolated dimer model for temperatures
above 20 K.

\begin{figure}[htbp]
\includegraphics[width=0.5\textwidth]{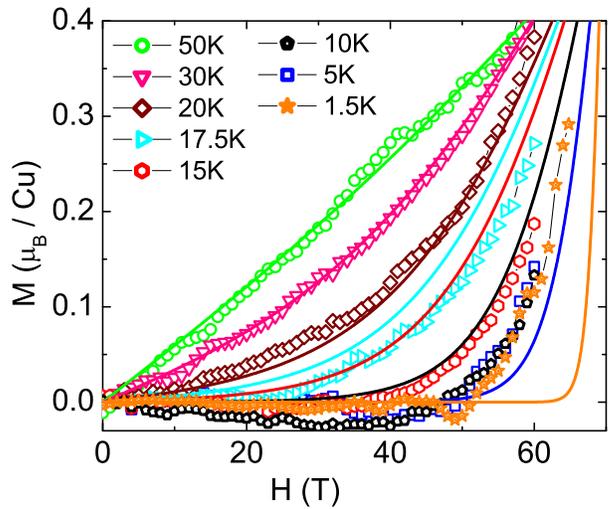}
\caption{Magnetization of Sr$_2$Cu(BO$_3$)$_2$ as a function of
applied field at different temperatures. Calculated values for
isolated dimer model with J = 100 K and g = 2.14 shown by solid
lines.}
\end{figure}

\begin{figure}[htbp]
\includegraphics[width=0.48\textwidth]{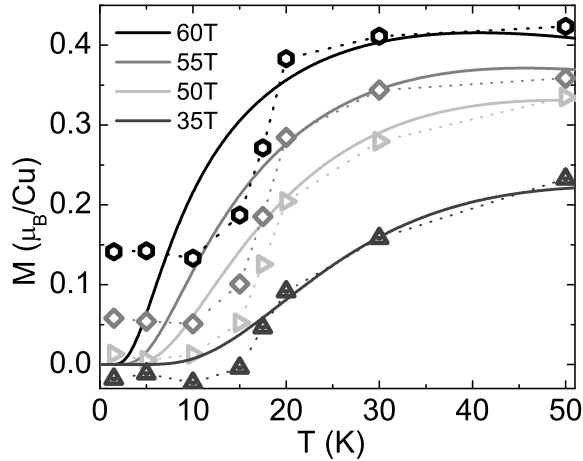}
\caption{Magnetization as a function of temperature for different
magnetic fields. Symbols (connected by dotted lines to guide the
eye) indicate measured magnetization data extracted from Figure 5,
solid lines indicate isolated dimer model with J = 100 K and g =
2.14}
\end{figure}

The observed high field magnetization behavior at low temperatures
indicates that Sr$_2$Cu(BO$_3$)$_2$ is not an isolated dimer
system. The upturn in magnetization due to the excitation of
singlet into triplet states occurs at a lower magnetic field H$_c$
$\sim$ 56 T than predicted by the isolated dimer model (67 - 74 T,
for g values between 2.2 and 2.0), indicating that the interdimer
exchange terms J' and J" have a significant effect. This effect
can be estimated if we assume that J' and J" are much smaller than
J and can be treated as a perturbation to the isolated
dimer model. These interdimer terms act on triplet
states and enable triplet hopping from first order in the
perturbation solution, resulting in a triplet dispersion relation
that reduces H$_c$ from J/g$\mu$$_B$. This is very different from
the Shastry Sutherland compound SrCu$_2$(BO$_3$)$_2$, for which
triplets are localized up to the fourth order in the perturbation
solution, due to frustration [4]. To obtain an estimate of H$_c$
in $\beta$-Sr$_2$Cu(BO$_3$)$_2$, the simplifying assumption can be
made that in the vicinity of level crossing, the eigenspace
comprises only the two degenerate singlet (S = 0) and lowest
triplet (S$^z$ = 1) states, which is a reasonable assumption given
the large value of J. These can be mapped on to an effective site which is empty (S = 0) or occupied (S$^z$ = 1) by a hardcore boson (equivalent to a pseudo-spin $\frac{1}{2}$.) The effective sites form a square lattice linked by J' and J" (Figure 2b.) Following the perturbative treatment of interdimer terms on a lattice of coupled spin $\frac{1}{2}$ chains [8,9] we obtain g$\mu${}$_B$H$_c$ = J - $\frac{J'+J"}{2}$ for the lattice in Figure 2b. 
For the observed range J/g$\mu${}$_B$ - H$_c$ $\simeq$ 11 - 18 T 
(depending on the value of g), we obtain $\frac{J'+J"}{2}$ $\sim$ 17 - 28 K. 
Experiments are in progress to directly measure the triplet dispersion 
and thereby independently determine the values of J' and J".

At low temperatures and in high magnetic fields, interactions
between the triplets will determine the nature of the ground
state, which is likely a canted antiferromagnet.  Within a picture
of delocalized triplets, and neglecting anisotropic terms 
such as Dzyaloshinkii Moriya exchange, 
it is natural to consider this as a Bose Einstein Condensate of 
hardcore bosons [9]. The temperature
independent finite magnetization below 20 K in high magnetic
fields (Figure 6) is presumably associated with this state,
following a similar temperature dependence to results from
numerical studies of weakly coupled dimer systems [10] and
experimental results for the related material TlCuCl$_3$ [11,12].
However, only a small portion of the phase diagram is accessible
due to the large value of the spin gap, and it is therefore
difficult to unambiguously identify the nature of the ground state
in high magnetic fields for this material.

In summary, we have made the first magnetization and heat capacity
measurements on single crystals of the new spin gap compound
$\beta$-Sr$_2$Cu(BO$_3$)$_2$. The material has a singlet groundstate comprising dimers with intradimer coupling J = 100 K. High field measurements reveal the presence of interdimer coupling, which results in the reduction of H$_c$ from J/g$\mu${}$_B$ and a finite low temperature magnetization at high fields. From these measurements, we estimate weak interdimer coupling $\frac{J'+J"}{2}$ $\sim$ 17 - 28 K. For fields above H$_c$, cooperative effects are expected to lead to an ordered ground state, the
nature of which remains to be determined.

We acknowledge helpful discussions with C. Batista. This work is supported by the National Science Foundation,
Division of Materials Research under DMR-0134613. Experiments
performed at the National High Magnetic Field Laboratory were
supported by the National Science Foundation through Cooperative
Grant No. DMR-9016241 and MRI Grant No. 0079641, the State of
Florida, and the Department of Energy. I.R.F acknowledges support
from the Alfred P. Sloan Foundation and S.E.S from the Mustard Seed Foundation.

\begin{acknowledgments}

\end{acknowledgments}

\end{document}